\documentclass[useAMS,usenatbib]{mn2e}
\usepackage{psfig,epsfig}
\usepackage{graphicx}
\begin{document}
\def\simlt{\mathrel{\rlap{\lower 3pt\hbox{$\sim$}}
        \raise 2.0pt\hbox{$<$}}}
\def\simgt{\mathrel{\rlap{\lower 3pt\hbox{$\sim$}}
        \raise 2.0pt\hbox{$>$}}}
\def\bj{b_{\rm\scriptscriptstyle J}}
\def\rt{r_{\rm\scriptscriptstyle T}}
\def\rp{r_{\rm\scriptscriptstyle P}}
\renewcommand{\labelenumi}{(\arabic{enumi})}
\title[MAMBO Observations at 240GHz of optically obscured {\it Spitzer} sources]
{MAMBO Observations at 240GHz of optically obscured {\it Spitzer} sources: source clumps and radio activity at high redshift}
\author[Paola Andreani et al.]
{\parbox[t]\textwidth{P. Andreani$^{1,2}$, M. Magliocchetti$^{3,4}$, G. De Zotti$^{5,4}$}\\
\tt $^1$ ESO, Karl-Schwarzschild-Str.2, D-85748, Garching, Germany\\
\tt $^2$ INAF, Osservatorio Astronomico di Trieste, Via Tiepolo 11, I-34143, Trieste, Italy\\
\tt $^3$ INAF, IFSI, via Fosso del Cavaliere 1100, Roma, Italy\\
\tt $^4$ SISSA, Via Beirut 4, I-34014, Trieste, Italy\\
\tt $^5$ INAF, Osservatorio Astronomico di Padova, Vicolo dell'Osservatorio 5,
I-35122 Padova, Italy\\} \maketitle
\begin{abstract}
Optically very faint ($R>25.5$) sources detected by the {\it Spitzer} Space
Telescope at 24$\mu$m represent a very interesting population at redshift
$z\sim[1.5-3]$. They exhibit strong clustering properties, implying that they
are hosted by very massive halos, and their mid-IR emission could be powered by
either dust-enshrouded star-formation and/or by an obscured AGN. We report
observations carried out with the MAMBO array at the IRAM 30m antenna on Pico
Veleta of a candidate protocluster with five optically obscured sources selected
from the 24$\mu$m {\it Spitzer} sample of the First Look Survey.
Interestingly, these sources appear to lie on
a high density filament aligned with the two radio jets of an AGN.
Four out of five of the observed sources were detected. We
combine these measurements with optical, infrared and radio observations to
 probe the nature of the candidate protocluster members. 
Our preliminary conclusions can be
summarized as follows: the Spectral Energy Distributions of all sources include both 
AGN and starburst contributions; the AGN contribution
to the bolometric luminosities ranges between 14 and 26\% of the total. Such a contribution 
is enough for the AGN to dominate the emission at 5.8, 8 and
24$\mu$m, while the stellar component, inferred from SED fitting, prevails at 1.25mm and at
$\lambda <4.5\mu$m. The present analysis suggests a coherent interplay at high-z between 
extended radio activity and the development of filamentary large-scale structures.
\end{abstract}


\section{Introduction}
\label{Intro}

Optically obscured 24$\mu$m {\it Spitzer} sources are among the most strongly
clustered objects in the Universe. Magliocchetti et al. (2007) studied a
complete sample of 793 optically faint ($R>25.5$) sources with $F_{24\mu \rm
m}\ge 0.35\,$mJy, drawn from the {\it Spitzer} Space Telescope  First Look
Survey (FLS) \citep[]{fadda06}. Their two-point correlation function has an
amplitude $A=(7\pm 2)\cdot 10^{-3}$ [$w(\theta)=A\theta^{-0.8}$, with $\theta$
in degrees], which is about a factor of 8 larger than that obtained for the
whole {\it Spitzer} $F_{24\mu\rm m}\ge 0.35$~mJy sample \citep[see Fig.~6
of][]{Mag07}. Their optical/mid-IR colours indicated that they are
ultraluminous far-infrared galaxies at $z \simeq 1.6-2.7$, a conclusion
supported by redshift determinations based on infrared spectroscopy (IRS; e.g.,
Yan et al. 2005, 2007; Houck et al. 2005; Weedman et al. 2006a, Desai et al., 2007)
and by comparisons with models for galaxy formation and evolution
\citep[e.g.,][]{Gra}. Their 3D comoving clustering radius was estimated to be
$r_0=[15.2^{+2.3}_{-2.6}]$~Mpc [$\xi(r)=(r/r_0)^{-1.8}$] (Magliocchetti et al.
2007).\hfill\break

 Clustering results obtained on Bootes and UKIDSS fields by
Brodwin et al. (2008) and Magliocchetti et al. (2008b) based on spectroscopic and photometric N(z)
converge, within the errors,  at giving similar high clustering lengths and halo masses (r$_0$ about
13 Mpc and masses 10$^{12.5}$-$10^{13}$ M$_\odot$ for S$>$0.4 mJy)\footnote{Note that the clustering length given
by Magliocchetti et al. (2008b) is in Mpc for $h=0.7$, not in $h^{-1}$
Mpc, as stated by Brodwin et al. (2008)}.
These measured clustering properties indicate that these sources are associated
to very massive dark matter (DM) halos, consistent with them
inhabiting the progenitors of groups/clusters of galaxies. Thus, these objects are
important tools to investigate the abundance of very massive halos at high $z$
and the evolution of large scale structure.

Bright, high-z QSOs, powered by the most massive black-holes (BHs), display
similar clustering properties \citep{Shen} and should therefore inhabit similar
halos at $z \sim 2$.
Obscured 24$\mu$m sources in the environment of bright
QSOs (and/or radiogalaxies) at $z \sim 2$ then likely trace the largest density
peaks and thus the very most massive DM halos. 
This could
be materialized by several filaments crossing at their position. High-$z$
radiogalaxies proved to be particularly effective
tracers of distant large scale overdensities that may collapse to form
present-day massive clusters of galaxies (Pentericci et al. 1997; Venemans
et al. 2007; Overzier et al. 2007; Miley \& De Breuck 2008). Since both
optically obscured $24\mu$m Spitzer sources and high-$z$ radiogalaxies are
preferentially found within the rare large-scale, large-amplitude peaks in
the primordial density fluctuation field, it is interesting to investigate
possible links between these two source populations.

A by-product of the clustering analysis by Magliocchetti et al. (2007) was the discovery
of several galaxy alignments, reminiscent of the
high-$z$ filaments found in numerical simulations which describe the evolution of
large-scale structure. In some cases, the candidate filaments were aligned with
radio jets, consistent with West's (1994) prediction that the radio axes
reflect the developing large-scale clustering pattern in the early Universe. If
so, these systems would be an important probe of the cosmological density field
at early epochs.

In this paper we start investigating the above issue by collecting all the available data for one
of these candidate filaments, and supplementing it 
with our own observations with the MAMBO array at the focus of the IRAM 30m antenna
at 1.2mm.

The addition of mm data is crucial in constraining the nature of the obscured 24$\mu$m sources.
In fact their nature is currently under debate: a large fraction of
them has radio spectral indices typical of {\it frustrated} radio-loud Active
Galactic Nuclei (AGNs) \citep{Mag08a} but their $q$ values -- i.e. the values of
the ratio between far-IR (FIR) and radio flux densities -- are similar to those
of star forming objects \citep[i.e.][]{Ibar}. Their colours indicate mixed AGN
and starburst properties (Magliocchetti et al. 2007; Pope et al 2008; Lonsdale et al. 2009).

FIR/mm observations are very effective for determining the bolometric output by measuring the redshifted peak of the dust
emission, for characterizing the broad band spectrum which allows to separate AGN from starburst emissions, and for studying the
parallel dust obscured evolution of super-massive BHs (SMBHs) and of their host galaxies. The AGN contribution produces
hotter dust and much higher $F$(mid-IR)$/F$(sub-mm) ratios than (sub)mm selected sources,  which are generally
high-$z$ starburst galaxies. Lutz et al. (2005) found a mean ratio $F_{1.2\rm mm}/F_{24\mu \rm m} = 0.15 \pm 0.03$ for a
sample of AGN-dominated optically obscured {\it Spitzer} FLS sources with $F_{24\mu \rm m}> 1\,$mJy. {\it
Spitzer} observations of starburst-dominated objects originally detected with SCUBA and MAMBO (Pope et al., 2006,
Ivison et al., 2007, Lonsdale et al., 2008) and with $F_{24\mu \rm m} >20\mu$Jy, instead have a median ${F_{850\mu
m}}/{F_{24\mu \rm m}}\simeq 25$ and ${F_{1.2\rm mm}}/{F_{24\mu \rm m}}\simeq 12$.

Estimating the AGN and star-forming components helps in investigating any
evolutionary connection between them, as the relationship between the
star-formation history and the growth of the active nucleus is a key ingredient
for understanding the physics of the evolution of both massive galaxies and
AGNs \citep{Gra,Scha}.

This work deals with a candidate
protocluster system of optically obscured 24$\mu$m {\it Spitzer} FLS sources
in the field of an extended radio-loud AGN. Such
a system: i) provides us with a sample of sources
with a 24$\mu$m flux distribution which is representative of that found for the whole
\cite{Mag07} dataset; ii) allows us to investigate in a direct way the
transition between 'pure' star-forming galaxies and AGN-dominated sources as
envisaged e.g. by the Granato et al. model; iii) perhaps most importantly, it
provides an excellent laboratory to study the large-scale interplay between AGN
activity (such as that produced by radio jets) and star formation in the
sources belonging to the protocluster.

The central source of this complex radio system is identified with
FLSVLA172010.2+592425.2 \cite{Condon} and at 8$^{\prime \prime}$ from it there
is another radio source, 7C1719+5927 \cite{Hales}. The flux densities
integrated over the whole system at 38, 330, 610 MHz, and 1.4GHz are of 2000,
143, 33.7, and 35.6 mJy, respectively. The 38, 330, and 1.4 GHz measurements
are consistent with a power-law spectrum ($S\propto \nu^{-\alpha}$, with
$\alpha \simeq 1$), while the 610 MHz flux density is a factor $\simeq 2$ below
that. The central source has an optical counterpart with a magnitude of $R= 23.67$ (Fadda et al.
2004). The spectroscopic redshift of this source is unknown.

Section 2 reports all the data  available on this sample and the new MAMBO
observations. Results of our analysis are given in \S\,3 and discussed in
\S\,4, while in \S\,5 we summarize our main conclusions.

\begin{figure*}
\includegraphics[width=16.0cm,height=12.cm]{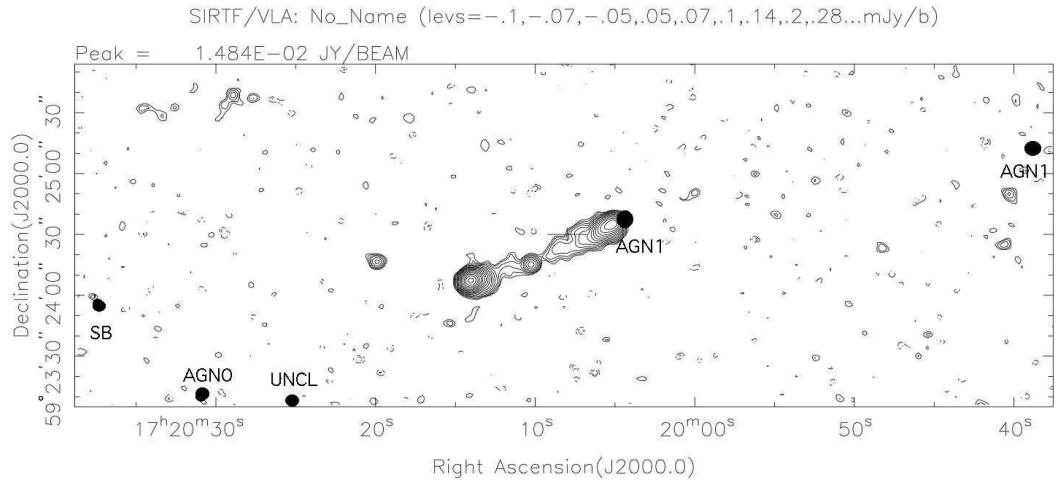}
\caption{{\it Spitzer}/VLA field centered at 17:20:05, +59:24:30. Filled circles correspond
to the FLS {\it Spitzer} 24$\mu$m sources we observed with MAMBO.}
\label{fig:field}
\end{figure*}

\section{Data}
\label{sample}

The source sample used in this work was drawn from a parent sample whose selection criteria
are outlined in this section.

\subsection{Sample selection and {\it Spitzer} data} \label{selection}

Of the original 793 $R>25.5$ sources in the Magliocchetti et al. (2007) sample,
510 have IRAC photometry. On the basis of these data, sources were classified
according to their $F_{8\mu\rm m}/F_{24\mu \rm m}$ and $F_{4.5\mu\rm
m}/F_{3.6\mu \rm m}$ colours \citep{Mag08a} as follows:

\begin{description}

\item[-] AGN0: $F_{24\mu \rm m}\ge 0.6\,$mJy;  $F_{8\mu\rm m}/F_{24\mu \rm m}\ge
0.1$;

\item[-] SB0 (starburst): $F_{24\mu \rm m}\ge 0.6\,$mJy;  $F_{8\mu\rm m}/F_{24\mu \rm m}<
0.1$;

 \item[-] AGN1: $F_{24\mu \rm m}< 0.6\,$mJy;  $F_{4.5\mu\rm m}/F_{3.6\mu \rm m}\ge
 4.5/3.6$;

\item[-] Unclassified: $F_{24\mu \rm m}< 0.6\,$mJy;  $F_{4.5\mu\rm m}/F_{3.6\mu \rm m}
< 4.5/3.6$ or no IRAC information. 

\end{description}

\noindent The above classification is  consistent with the results of
mid-infrared (IRS) spectroscopy of samples of heavily obscured 24$\mu$m-bright objects
(brighter than $\sim 0.7$--1~mJy). The
majority of them have spectral shapes indicative of a buried AGN
(Martinez-Sansigre et al. 2006; Yan et al. 2005, 2007; Brand et al. 2007;
Weedman et al. 2006a,b; Pope et al. 2008).


\begin{table*}
\begin{center}
\begin{tabular}{cccccccccc}
\hline
\hline
\\
\multicolumn{1}{c}{{\rm Source name}} &
\multicolumn{1}{c}{{\rm R$_{\rm mag}$}} &
\multicolumn{1}{c}{{\rm 3.6$\mu$m}} &
\multicolumn{1}{c}{{\rm 4.5$\mu$m}} &
\multicolumn{1}{c}{{\rm 5.8$\mu$m}} &
\multicolumn{1}{c}{{\rm 8$\mu$m}} &
\multicolumn{1}{c}{{\rm 24$\mu$m}} &
\multicolumn{1}{c}{{\rm 1.4GHz}} &
\multicolumn{1}{c}{{\rm 610MHz}} &
\multicolumn{1}{c}{{\rm ID}}  \\
\multicolumn{1}{c}{{\rm }} &
\multicolumn{1}{c}{{\rm }} &
\multicolumn{1}{c}{{\rm $\mu$Jy}} &
\multicolumn{1}{c}{{\rm $\mu$Jy}} &
\multicolumn{1}{c}{{\rm $\mu$Jy}} &
\multicolumn{1}{c}{{\rm $\mu$Jy}} &
\multicolumn{1}{c}{{\rm $\mu$Jy}} &
\multicolumn{1}{c}{{\rm $\mu$Jy}} &
\multicolumn{1}{c}{{\rm $\mu$Jy}} &
\multicolumn{1}{c}{{\rm }} \\
\hline
\\
J172037.75+592359.3 & $>$25.5 & 29.93$\pm$4.33 & 45.84$\pm$6.24 & $<$100 & $<$100         & 400$\pm$80 & $<$100     & $<$182 & SB \\
J172030.92+592308.3 & $>$25.5 & 64.68$\pm$7.84 & 53.51$\pm$7.15 & $<$100 & 191.4$\pm$24.3 & 390$\pm$75 & $<$100     & $<$182 & AGN0\\
J172005.01+592430.0 & $>$25.5 & 24.36$\pm$3.65 & 33.52$\pm$5.18 & $<$100 & $<$100         & 440$\pm$88 &  --        & --     & AGN1\\
J171939.17+592512.4 & $>$25.5 & 20.89$\pm$3.37 & 36.59$\pm$5.44 & $<$100 & $<$100         & 380$\pm$74 & 157$\pm$27 & $<$182 & AGN1\\
J172025.20+592308.0 & $>$25.5 & $<$100         & $<$100         & $<$100 & $<$100         & 350$\pm$70 & $<$100     & $<$182 & UNCL\\
\\
\hline
\hline
\\
\end{tabular}
\caption{Optically obscured 24$\mu$m {\it Spitzer} sources observed with MAMBO
during Winter 2008. The table lists the optical and {\it Spitzer} photometry
available for these sources. Last two columns report the radio measurements
when available. The source J172005+592430 lies on top of a lobe of the radio
source (see Fig.~\protect\ref{fig:field}), that hides its possible radio emission.
Upper limits are at the $3\sigma$ level. The last column shows the type of identification assigned by Magliocchetti et al. (2007) to
these sources. We have assigned in our analysis an error-bar of 20\% to the 24$\mu$m fluxes (cfr text).}\label{TabAll}
\end{center}
\end{table*}

\begin{table*}
\begin{center}
\begin{tabular}{cccccccccc}
\hline
\hline
\\
\multicolumn{1}{c}{{\rm Source name}} &
\multicolumn{1}{c}{{\rm weighted}} &
\multicolumn{1}{c}{{\rm first}} &
\multicolumn{1}{c}{{\rm atmospheric}} &
\multicolumn{1}{c}{{\rm average}} &
\multicolumn{1}{c}{{\rm on source}} &
\multicolumn{1}{c}{{\rm second}} &
\multicolumn{1}{c}{{\rm atmospheric}} &
\multicolumn{1}{c}{{\rm average}} &
\multicolumn{1}{c}{{\rm on source}} \\
\multicolumn{1}{c}{} &
\multicolumn{1}{c}{{\rm flux}} &
\multicolumn{1}{c}{{\rm observation}} &
\multicolumn{1}{c}{{\rm transmission}} &
\multicolumn{1}{c}{{\rm atm noise}} &
\multicolumn{1}{c}{{\rm int time}} &
\multicolumn{1}{c}{{\rm observation}} &
\multicolumn{1}{c}{{\rm transmission}} &
\multicolumn{1}{c}{{\rm atm noise}} &
\multicolumn{1}{c}{{\rm int time}} \\
\multicolumn{1}{c}{{\rm }} &
\multicolumn{1}{c}{{\rm (mJy)}} &
\multicolumn{1}{c}{{\rm (mJy)}}&
\multicolumn{1}{c}{{$\tau$}} &
\multicolumn{1}{c}{{\rm (mJy$s^{1/2}$)}} &
\multicolumn{1}{c}{{\rm (min) }} &
\multicolumn{1}{c}{{\rm (mJy)}} &
\multicolumn{1}{c}{{$\tau$}} &
\multicolumn{1}{c}{{\rm (mJy$s^{1/2}$)}} &
\multicolumn{1}{c}{{\rm (min) }} \\
\\
\hline
J172037.75+592359.3 &  1.4$\pm$0.5 & 1.2$\pm$0.5   & 0.35 & 36 & 160 & 3.4$\pm$1.4   & 0.15 & 90 & 52 \\
J172030.92+592308.3 &       $<$1.2 & 0.12$\pm$0.63 & 0.16 & 50 &  48 &-0.13$\pm$0.58 & 0.36 & 40  & 64\\
J172005.01+592430.0 &  1.6$\pm$0.3 & 1.0$\pm$0.5   & 0.37 & 46 &  80 & 2.2$\pm$0.5   & 0.27 & 63  & 72 \\
J171939.17+592512.4 &  1.4$\pm$0.5 & 1.3$\pm$0.6   & 0.5  & 90 & 102 & 1.3$\pm$0.6   & 0.3  & 40  & 52 \\
J172025.20+592308.0 &  1.8$\pm$0.4$^\dagger$ & 0.7$\pm$0.6   & 0.35 & 40 &  60 & 2.5$\pm$0.6   & 0.26 & 60  & 80\\
\hline \hline
\\
\end{tabular}
\caption{MAMBO observations of 24~$\mu$m optically obscured {\it Spitzer}
sources. The sources observed were previously classified -- according to our
colour selection criterion (Magliocchetti et al., 2007) -- as either starbursts or obscured
AGNs. The only undetected source was indeed classified as an AGN and its flux
upper limit is 1.2~mJy (3$\sigma$). Atmospheric values in columns 4,5,8,9 refer to average values during the observations.
Errors on the fluxes are statistical. Final flux is computed as a weighted average.\hfill\break
\noindent
$\dagger$ Because of the discrepancy between the two measurements of this source, we assign in our analysis
an additional systematic error-bar of 20\% to its mm flux.}\label{TabObsLog}
\end{center}
\end{table*}

\subsection{Radio observations: selection of protoclusters}
Magliocchetti et al. (2008a) searched for radio counterparts to optically
obscured {\it Spitzer} sources by cross-correlating the source sample of
Magliocchetti et al. (2007) with the radio source catalogs by Condon et al.
(2003) at 1.4GHz, Morganti et al. (2004) at 1.4GHz and Garn et al. (2007) at 610 MHz.

A number of the selected sources turned out to lie in overdense regions of the sky
in the neighborhood of radio-loud AGNs (see, e.g., Fig.~\ref{fig:field}). To
pursue our research further we then selected small complete subsamples in those
high density regions with deep radio imaging and hosting a radio-loud AGN. In
this paper we concentrate on the field shown in Fig.~\ref{fig:field}.
The available {\it Spitzer}, radio and optical
photometric data are listed in Table \ref{TabAll}.

\begin{figure*}
\includegraphics[width=14.0cm]{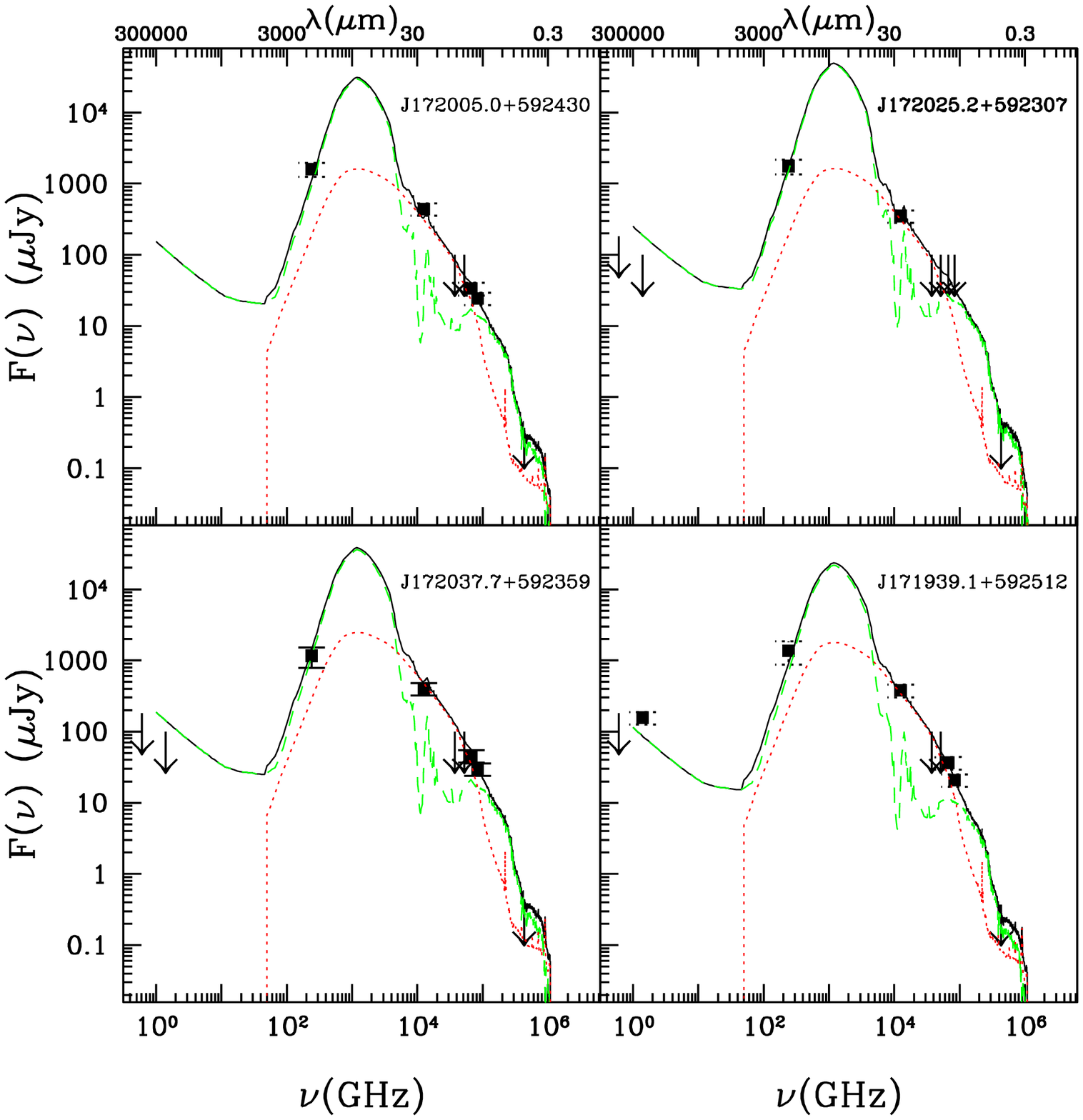}
\caption{Spectral energy distributions of the four sources detected with MAMBO in our
sample. Photometric points from the R magnitude down to the radio fluxes at 1.4
GHz and the 610 MHz are shown together with two spectral energy distributions which best
fit the observed points: a starburst component represented by Arp220 SED (green dashed line) and an obscured
torus (Polletta et al., 2007, red dotted line). The SEDs are shifted to a redshift of 1.73. The corresponding bolometric
luminosities are 5-7 $\times$10$^{12}$ L$_\odot$ (see Table~\ref{TabSED}).\label{figSEDs}}
\end{figure*}

\subsection{Redshift constraints: IRAC and optical observations}

IRAC photometry can also be used to provide rough redshift estimates for those
sources whose SEDs present a contribution from an evolved stellar population in
the 3.6$\mu$m and 4.5$\mu$m bands. Based on the form of the SED in this
wavelength range \citep[see Fig.~6 in][]{Mag08a}, and in particular on its
interpolated slope $\beta$ ($F\propto\lambda^{\beta}$) between 3.6$\mu$m and
4.5$\mu$m, we estimated that the redshifts of these objects are around $z\sim
1.7$. This makes it plausible that they are physically associated and not just close to one another
due to projection effects.

An independent assessment of the redshift range for this source population comes from
the measured spectroscopic redshifts of optically faint 24$\mu$m {\it Spitzer}
sources by Dey et al. (2008) 
using the IRS {\it Spitzer} spectrograph and Keck optical spectroscopy. If we
adopt Dey et al.'s colour criterion, our sources have a median redshift
$<z>=1.8$--1.9, roughly consistent with our previous estimate.

\subsection{MAMBO observations of the candidate protocluster}

Data were taken with the 117-channel Max Planck Millimetre Bolometer Array
(MAMBO; Kreysa et al. 1998) at the focus of the 30m IRAM antenna at Pico
Veleta. Chopping photometry at 2\,Hz with 32$^{\prime\prime}$ throw in azimuth
was gathered at 250 GHz (1.25 mm) in observing pool during Winter 2008 towards
the sources listed in Tables ~\ref{TabAll} and ~\ref{TabObsLog}. Data were obtained in good
observing conditions (250 GHz atmospheric opacity better than 0.2 and low sky
noise), as resulting from the regular checks of atmospheric transmission and
noise. Pointing was checked every hour with a nearby quasar. Each source has
been observed in at least two independent runs. Data were reduced with the
MOPSIC software \cite{Zylka}.
\hfill\break
To check for any malfunctioning of the detecting system with consequent
instabilities of the bolometer output, we inspected subscan by subscan and
reduced the data separately for each observing chunk. Subscans containing
spikes (less than 10\% of the total) were discarded. Observing logs with total
integration times on sources, atmospheric parameters, fluxes at each run and the final calibrated
fluxes with statistical errorbars are reported in Table \ref{TabObsLog}.
Longer integration times were necessary because of the higher sky noise during
some of the observations. Statistically, the sky noise was behaving properly as it scales
with the square root of the integration time.

\begin{figure}
\includegraphics[width=8.0cm]{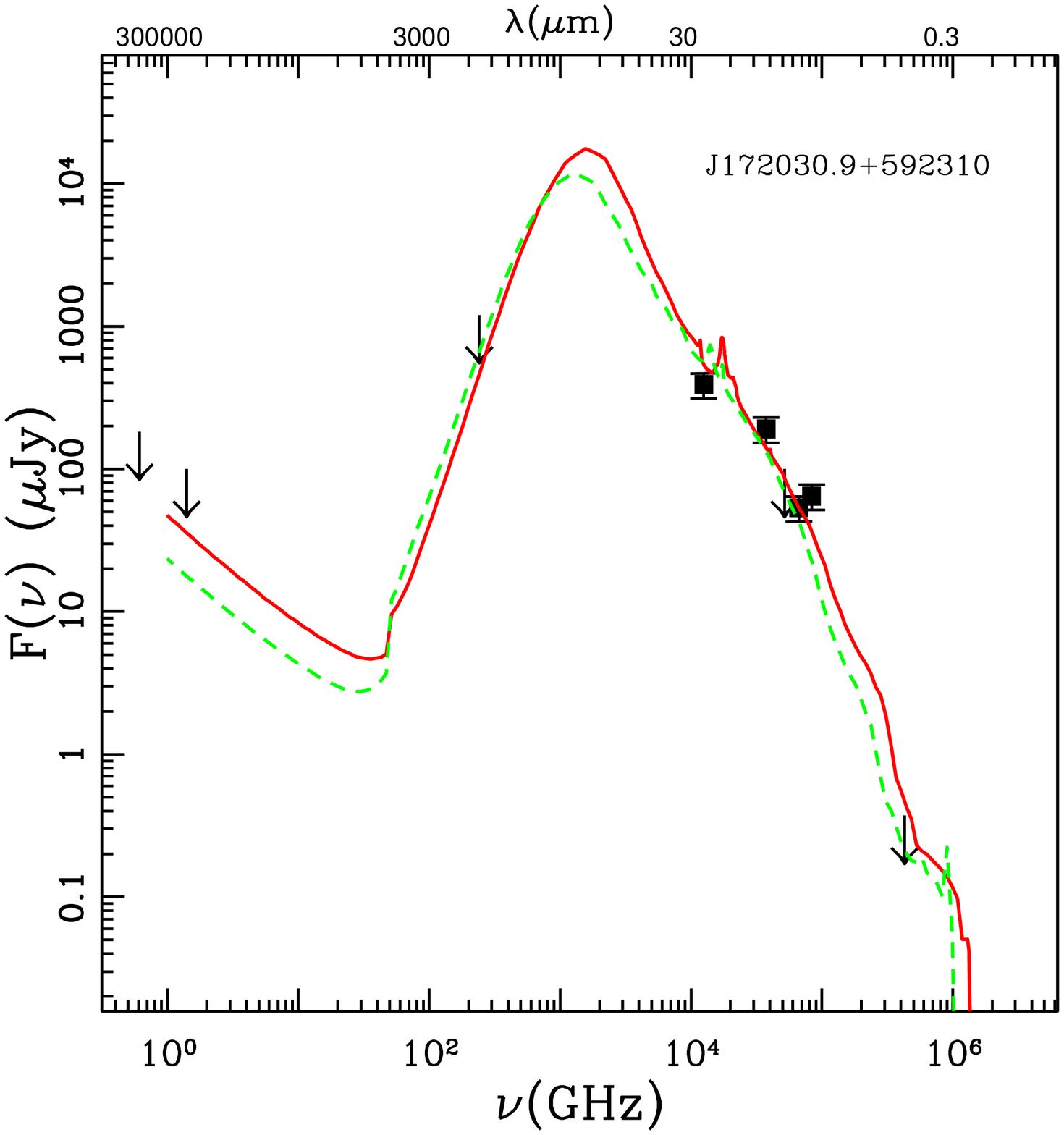}
\caption{Spectral energy distribution as in Figure~\ref{figSEDs} for the AGN0.
The green dashed line shows the best fit, obtained adding the
SED of Arp~220 with Polletta's ``obscured torus'' template, red-shifted to $z=1.214$;
the red solid line shows the best-fit (giving a much worse $\chi^2$) derived with the same
combination of SEDs red-shifted to $z=1.73$. We classify this source as a type 2 AGN candidate,
i.e. an obscured AGN.\label{figSEDAGN}}
\end{figure}

\section{Results}
\label{results}

\subsection{Interpreting the observed spectral energy distributions}

Figure~\ref{figSEDs} shows the photometric points from the R magnitude down to
the radio fluxes at 1.4 GHz and 610 MHz for the sources of our sample.
This also includes the MAMBO observations presented in Section 2.4.
The original 24$\mu$m errorbars given by Fadda et al (2006) for these sources are
unrealistically small (1.5\%); on the suggestion by the referee, we have adopted 20\%
error bars, as estimated for similar (80s) 
integration times.

In Table \ref{TabSED} we report the results from
two different types of fits and the corresponding
$\chi ^2$ values and luminosities.  The first template used is the Arp220 SED, while the second
has two components: Arp220 SED and an obscured torus template made available by Polletta
et al. (2007). In this latter case we could estimate the contribution to the bolometric luminosity
of the two components.
We find that the Arp~220 Spectral Energy Distribution (SED) provides the best single component
fit for all the observed SEDs.  In fact, all of the 30 SEDs of nearby star-forming galaxies presented by Vega et al.
(2008) and the other template SEDs made available by
Polletta (Polletta et al. 2007) give substantially worse fits.
By then using the Arp~220 SED as 'bona fide' template, we get a minimum
$\chi^2$ fit at the same value of the redshift, $z=1.73$, for each object belonging to the filament
separately, with the exception of J172030.9+592310 which is classified as
an AGN0 from IRAC photometry (see Table 1).
At this redshift,
the sources have bolometric luminosities in the range 5--$7\cdot 10^{12}$
L$_\odot$ and star formation rates of $\sim 1000$ M$_\odot$/yr (see Table \ref{TabSED})\footnote{The star formation rate
was estimated from the total far-IR luminosity, which is essentially equal to the bolometric luminosity,
using eq. (4) of Kennicutt (1998).}.
Obviously, this surmise can only be confirmed by spectroscopic redshift measurements.

No really good fit could be found for J172030.9+592310. In this case, neither
the Arp~220 template, nor any other template mentioned above, can be accepted as
a realistic description of its SED. The lowest $\chi^2$, still quite unsatisfactory
($\chi^2=23.9$), fit was obtained by adding to the SED of Arp 220 the
Polletta's ``torus'' template (heavily obscured AGN) at $z=1.214$ (green line
in Fig.~\ref{figSEDAGN}). Additionally, Figure ~\ref{figSEDAGN} shows as a red line the
best-fit with the same combination of SEDs red-shifted to $z=1.73$. This shift
in $z$ substantially worsens the $\chi^2$ value ($\chi^2 = 48.5$).

In Table \ref{TabSED} we give the estimated bolometric luminosities af the sources for the two models we have considered, i.e.
pure starburst SED (I fit) and starburst plus AGN  (II fit). The errors on the total luminosities were computed finding the
contours in the parameter space for which $\chi^2 = \chi^2_{\rm min}+1$ and computing the maximum and minimum luminosities
on these contours. The resulting errors are of 30--35\%. The main contribution to them comes from the uncertainties on redshifts
of the sources, which turn out to be $\Delta z \simeq $0.17--0.20.  In the case of the SB$+$AGN fit we give the estimated
bolometric luminosities of each component. The global bolometric
luminosity is therefore the sum of the two contributions, and it is used to compute, in a consistent way, the SB and AGN fractions.
The estimate of the bolometric luminosity clearly depends on the used SED template. No surprise that 2 substantially different
 templates yield different results. Note, however, that differences are not very large. In the worst case (source J171939.1,
leaving aside the pathological second source), going from a pure Arp220 template to a SB+AGN template decreases the bolometric
 luminosity by 36\%. We do not attach any profound implication to the fact that for 3 sources the bolometric luminosity decreases
 from case I to case II, while only in one case (last source) it increases: this is most likely happening by chance. 

\begin{table*}
\begin{center}
\begin{tabular}{cccccc}
\hline
\hline
\\
\multicolumn{1}{c}{{\rm Source name}} &
\multicolumn{1}{c}{{\rm $\chi^2$ I fit}} &
\multicolumn{1}{c}{{\rm L$_{\rm bol}$ I fit}} &
\multicolumn{1}{c}{{\rm $\chi^2$ II fit}} &
\multicolumn{1}{c}{{\rm L$_{\rm SB}$ II fit}} &
\multicolumn{1}{c}{{\rm L$_{\rm AGN}$ II fit}} \\
\multicolumn{1}{c}{} &
\multicolumn{1}{c}{(5 d.o.f.)} &
\multicolumn{1}{c}{{\rm L$_\odot$}} &
\multicolumn{1}{c}{(4 d.o.f.) } &
\multicolumn{1}{c}{{\rm L$_{\odot}$}} &
\multicolumn{1}{c}{{\rm L$_{\odot}$}} \\
\\
\hline
J172037.7+592359 &  6.6  & $6.5\times 10^{12}$  & 0.23 & 4.1$\times 10^{12}$ &1.2$\times 10^{12}$ \\
J172030.9+592310 &  81   & $7.1\times 10^{12}$  & 23.9 & 1.0$\times 10^{12}$ &0.5$\times 10^{12}$  \\
J172005.0+592430 &  2.4  & $5.7\times 10^{12}$  & 2.4   & 3.4$\times 10^{12}$ &0.8$\times 10^{12}$ \\
J171939.1+592512 &  4.6  & $5.3\times 10^{12}$  & 1.4   & 2.5$\times 10^{12}$ &0.9$\times 10^{12}$  \\
J172025.2+592307 &  2.5  & $5.2\times 10^{12}$  & 1.1   & 5.5$\times 10^{12}$ &0.8$\times 10^{12}$ \\

\hline \hline
\\
\end{tabular}
\caption{Results of the minimum $\chi2$ fits. Fit I is obtained adopting the Arp 220 SED (columns 2 and 3). Adding an
obscured AGN component (fit II, columns 4-6) improves the $\chi^2$ values for three of the sources (we leave aside
the source J172030.9+592310 for which no acceptable fit was obtained; see text). We note that these results are
only indicative as the number of photometric points is too small to allow
any firm conclusion. The errorbars of the 24$\mu$m fluxes have been set to
20\%. Errorbars on the total luminosities are estimated to be 30\% - 35\% (see text for detail).}\label{TabSED}
\end{center}
\end{table*}

Based on the SED analysis which is listed in Table 3, we find that the nature of our candidate protocluster members can
be summarized as follows: the AGN contribution to the
bolometric luminosities ranges between 14 and 26\% of the total. A $\simeq 14\%$ contribution to the total bolometric
luminosity is enough for the AGN to dominate the emission at 5.8, 8 and 24$\mu$m, while the SB dominates at 1.25mm and at
$\lambda <4.5\mu$m. The above fits are consistent with all these sources being at the same redshift.

\begin{figure}
\includegraphics[width=8.0cm]{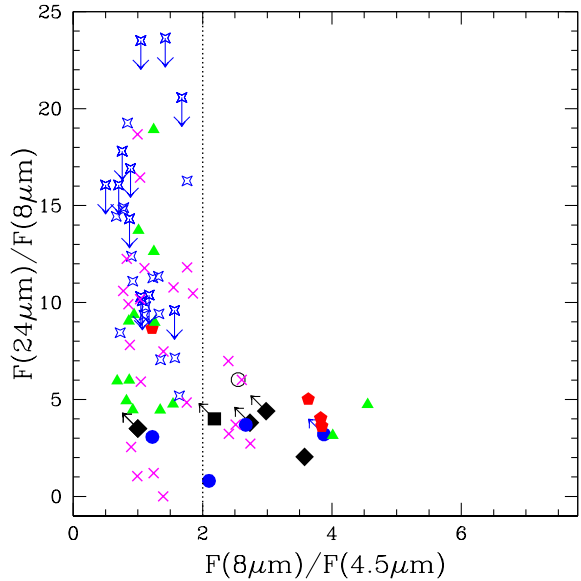}
\caption{The colour--colour $F({24\mu\hbox{m}})/F({8\mu\hbox{m}})$ vs.
$F({8\mu\hbox{m}})/F({4.5\mu\hbox{m}})$ plot is used as a diagnostic tool to discriminate
between starburst and AGN dominated SEDs. The vertical dashed line corresponds
to the ratio ($F({8\mu\hbox{m}})/F({4.5\mu\hbox{m}})=2$, proposed by Pope et al. (2008)
as the boundary between SB (having ratios $<2$) and AGN dominated SEDs. Filled squares correspond
to the {\it Spitzer} sources presented in this work (we note that for all but one case the 8$\mu$m measurement is a mere upper limit);
filled triangles: Ivison et al. (2004); filled pentagon: Younger et al. (2007); filled circles: Sawicki \& Webb (2005);
crosses: Pope at. (2006); four-points stars: Lonsdale et al. (2008).}
\label{colcol1}
\end{figure}


\subsection{Nature of the obscured {\it Spitzer} 24$\mu$m sources}

The information provided by the spectral energy distribution is not, however, compelling
and the nature of these sources cannot be entirely evinced. However
colour--colour plots may be further exploited to support our previous findings on the origin of
the source emission. We compare in this section the colours of the sources
presented in this work with those of other 24$\mu$m {\it Spitzer} sources,
extracted with different colour criteria, and with those of the SMGs, in the
effort to elucidate the nature of the objects belonging to our candidate protocluster.

\subsubsection{Colour--colour diagrammes}

In the original paper by Magliocchetti et al. (2007) the distribution of the $F_{8\,\mu \rm m}/F_{24\,\mu \rm m}$ 
versus the $F_{\rm R}/F_{24\,\mu \rm m}$ colours has been shown as a powerful tool to discriminate
between starburst dominated ($F_{8\,\mu \rm m}/F_{24\,\mu \rm m}<$1)
from AGN dominated ($F_{8\,\mu \rm m}/F_{24\,\mu \rm m}>$1) objects among the optical obscured 24$\mu$m sources
(see their Figure 2).
The SED fitting results obtained in section 3.1 for all sources but J172030.9+592310
are consistent with the preliminary estimates obtained
from the location of such sources in the [$\nu_{24} F(24\mu\hbox{m})/\nu _{\rm R}F_{\rm R}$,
$\nu_{24} F(24\mu\hbox{m})/\nu_8 F(8\mu$m)] colour--colour plane. Sources of this sample have
$\log (\nu _{24} F(24\mu\hbox{m})/(\nu_R F_{\rm R}))>1.74$
and $\log(\nu _{24} F(24\mu$m)$/\nu _{8} F(8\mu$m$))>0.55$,
corresponding to the colours of NGC 4418 (heavily obscured galaxy) shifted to a
redshift of 2 \cite{Lu05}, with the only exception of the AGN0 which has
$\log (\nu_{24} F(24\mu\hbox{m})/\nu_8 F(8\mu m))=0.3$ and
$\log (\nu _{24} F(24\mu\hbox{m})/(\nu_R F_{\rm R}))>1.74$, putting it
closer to the expected colours of the NGC1068 nucleus at a redshift around 2.

The values of the 1.2mm/$24\mu$m flux ratios which are higher for the sources of the {\it
Spitzer} sample of this work range from 3 to 5, except for the AGN0, which has
$F(1.2\hbox{mm})/F(24\mu\hbox{m})<3$. On a $F(24\mu\hbox{m})$ versus redshift
plot these values coincide with those obtained by Lonsdale et al. (2008) for
SWIRE sources (see their Fig.~8). Lonsdale et al. (2008) have observed a large
part of the SWIRE sample at 1.2mm with the MAMBO array and have shown that the
SWIRE sources and sub-mm selected galaxies (SMGs) have similar mm fluxes, but
the latter have $F(1.2\hbox{mm})/F(24\mu\hbox{m})$ ratios higher than the
former by a factor of 2 to 10. {\it Spitzer} optically obscured sources have
values intermediate between those of the SMGs and those characteristic of AGNs.
This finding is independently confirmed by IRS spectra of SWIRE sources which
show that part of the 24$\mu$m flux is due to a hidden AGN.

In Fig.~\ref{colcol1} the location in the [$F({24\mu\hbox{m}})/F({8\mu\hbox{m}})$,
$F({8\mu\hbox{m}})/F({4.5\mu\hbox{m}})$] plane of our four sources with IRAC
fluxes are compared with those of the SMGs and of the dust obscured galaxies
(DOGs; Pope et al. 2008). This colour--colour diagram was used by Ivison et al.
(2004) as a diagnostic tool to discriminate between starburst (SB) dominated
and AGN dominated SMGs detected by {\it Spitzer}. Their SMGs turned out to lie
in the locus of SB dominated galaxies, except for two with AGN colours. The same
diagram was used by Pope at al. (2008) to classify their DOGs ($R-[24]>14$ and
$F({24\mu\hbox{m}}) >100\,\mu$Jy). These authors proposed the ratio
$F({8\mu\hbox{m}})/F({4.5\mu\hbox{m}})=2$ as the boundary between SB (lower
values of the ratio) and AGN dominated SEDs. The conclusion was that 80\% of
the DOGs are starburst dominated.

All the sources analysed in this work but one (cfr Table 1 and Figures ~\ref{figSEDs}, ~\ref{figSEDAGN})
have 8$\mu\hbox{m}$ fluxes below the sensitivity threshold of the relevant IRAC channel (i.e. 100 $\mu$Jy),
yielding upper limits to the ${8\mu\hbox{m}}/{4.5\mu\hbox{m}}$ ratio in the
range 2.2--3, and lower limits to the ${24\mu\hbox{m}}/{8\mu\hbox{m}}$ ratio in
the range 3.8--4.4.
For the measured 24$\mu\hbox{m}$ fluxes, this is indicative of a starforming SED as an
AGN/power-law SED most likely would have implied detectable 8$\mu\hbox{m}$ fluxes.
In addition, adopting the Pope et al.'s (2008) criterion, our sources may
have a SB component, though the 8$\mu$m observations are not deep enough to certain.
The only exception is J172030.9+592310 which has ${8\mu\hbox{m}}/{4.5\mu\hbox{m}}\simeq 3.6$ and is
therefore clearly in the realm of AGNs.

IRAC photometry helps further in diagnosing the nature of these sources. In
particular the [${8\mu\hbox{m}}/{4.5\mu\hbox{m}}$,
${5.8\mu\hbox{m}}/{3.6\mu\hbox{m}}$] colours can be used to distinguish SEDs
with evolved stellar population (which would determine a 'bump' in the SED between
3 and 8$\mu$m in the observed frame) and those with a power-law
continuum all through the IRAC channels, characteristic of AGNs (i.e.
Magliocchetti et al., 2008a and Yun et al. 2008).
Although we cannot draw
any strong conclusion because of the lack of accurate IRAC photometry, we have
at least upper limits on the flux ratios for the four sources detected by
MAMBO, in the ranges $\log F({8\mu\hbox{m}})/F({4.5\mu\hbox{m}}) \le 0.3$--0.5
and $\log F({5.8\mu\hbox{m}})/F({3.6\mu\hbox{m}}) \le 0.5$--0.7, while the AGN0
has $\log F({8\mu\hbox{m}})/F({4.5\mu\hbox{m}}) = 0.55 $ and $\log
F({5.8\mu\hbox{m}})/F({3.6\mu\hbox{m}}) \le 0.2$. The diagnostic diagrams of
Yun et al. (2008) then confirm the conclusions above, i.e. the sources studied in this
work are dominated by star formation (SF) processes and have similar properties to the DOGs (Pope et al. 2008).

As we have just seen, {\it Spitzer} data give strong hints on the nature of our
sources. However, they do not allow us to unambiguously characterize them as
they cannot provide direct information on their relationship with SMGs and on the importance
of the AGN contribution to their SEDs.\\
We can then go a step forward and exploit further information.
For example, by using the mm and radio fluxes, together with the upper limits, we investigate
the well-known, tight relationship
between radio and far-IR emissions. The origin of this relation is attributed
to star formation processes and has been extensively investigated for a
quarter of a century. It is firmly established for the local Universe (i.e.
Helou, Soifer, Rowan-Robinson 1985) and appears to also extend to higher redshifts
\citep[e.g.][]{Garrett,Appleton,Ibar,GarnAlexander}.

By then comparing the [${850\mu\hbox{m}}/{1.4\hbox{GHz}}$,
${850\mu\hbox{m}}/{24\mu\hbox{m}}$] colours of the {\it Spitzer} obscured sources
with those SMGs having a radio and {\it Spitzer} counterpart we find that the latter
have lower ${850\mu\hbox{m}}/{1.4\hbox{GHz}}$ and ${850\mu\hbox{m}}/{24\mu\hbox{m}}$
ratios but both follow very tightly the correlation between the two colours, indicating once again
that the sources in our sample with measured MAMBO fluxes are dominated by SF processes.

The above correlation can also be exploited to estimate the redshift from the
$F({850\mu\hbox{m}})/F({1.4\hbox{GHz}})$ ratio, using the Carilli \& Yun
(1999) plot. We convert the MAMBO 1.2 mm fluxes to $850\mu$m fluxes using the median
ratio $F_{850\mu\rm m}/F_{1.2\rm mm}\simeq 2.5$, as found by Greve et al.
(2004), for SCUBA galaxies at $z\sim 2$.
The source with 1.4 GHz radio flux has
${850\mu\hbox{m}}/{1.4\hbox{GHz}} \simeq 22$, placing it in the redshift
interval 1--2.2. The 2 sources with radio upper limits have
${850\mu\hbox{m}}/{1.4\hbox{GHz}} > 30$ and $>45$ and redshifts $>1.3$ and
$>1.5$, respectively.\hfill\break

As for the Spitzer source coinciding with the
radio lobe, the convolution of the radio map with a Gaussian of FWHM
corresponding to the 24$\mu$m MIPS beam (6$^{\prime\prime}$) gives a flux
at 1.4 GHz of $\sim$ 3 mJy. The MIPS beam is comparable to the resolution
of the radio VLA 1.4 GHz observations, which was 5$^{\prime \prime}$. 
\hfill\break
This value could be entirely associated to the lobe itself or may be due to
the overlap of two components: one associated to the 24$\mu$m source and one to the
lobe. We therefore consider this value as un upper limit to the real radio emission of this
Spitzer source.
In any case the correlation FIR-radio, which is the underlying assumption
to infer the redshift, cannot be applied.

\section{Discussion}
\label{discussion}

As is widely known, the flux of sources at a redshift around 2 in the {\it
Spitzer} 24$\mu$m channel may contain emission from the PAH feature at a rest
frame wavelength of 7.7$\mu$m. Since this feature is associated to star forming
processes, the sample selection based on the 24$\mu$m flux is biased as it
enhances the probability of detecting star forming sources at this epoch.

This flux band has been extensively used to estimate star formation in galaxies
(see, e.g., Calzetti 2008 for a recent review and Rieke et al. 2009 for local
galaxies). However, the use of this band alone requires many {\it a priori}
assumptions (e.g. comparison with low-redshift SEDs, PAH
evolution, metallicity and the conditions in the photon dissociation regions)
on the overall spectral energy distribution and completely misses star
formation processes associated to molecular gas. We have estimated the star
formation rate by fitting all the available photometric data on these
sources. It follows that the uncertainties in this case are more related to the unavailability
of the far-infrared fluxes, which can only be retrieved with future Herschel
observations, and to the scant knowledge of the AGN contribution in the mid-IR
domain. Because of their optical faintness ($R>$25.5) spectroscopic redshifts of
these objects are hard to obtain and they will be ideal candidates for ALMA follow-up
observations.

Given the above difficulties, we tried to infer the nature of the sources
in our sample from colour--colour plots, coming
to the conclusion that the {\it Spitzer} optically obscured sources belonging to
the analysed protocluster are likely to host
an AGN and their mid-IR fluxes may be affected by its emission. Quantifying
this effect will be only possible when far-IR observations will allow us to
derive the starburst contribution to the SEDs.

One of the sources studied in this work has its position coinciding with one of
the radio lobes of the radio loud AGN. At the locations of the radio core and
of the second radio lobe there are no {\it Spitzer} detected sources. The
convolution of the radio flux coinciding with the location of the {\it Spitzer}
source gives a flux at 1.4GHz of $\sim$ 3 mJy, which we consider as un upper limit
to its real radio emission.

The central source of this complex radio system is identified with
FLSVLA172010.2+592425.2 \cite{Condon} and at 8$^{\prime \prime}$ from it there
is another radio source, 7C1719+5927 \cite{Hales}. The flux densities
integrated over the whole system at 38, 330, 610 MHz, and 1.4GHz are of 2000,
143, 33.7, and 35.6 mJy, respectively. The 38, 330, and 1.4 GHz measurements
are consistent with a power-law spectrum ($S\propto \nu^{-\alpha}$, with
$\alpha \simeq 1$), while the 610 MHz flux density is a factor $\simeq 2$ below
that. This atypical behaviour deserves further investigation. The central
source has an optical counterpart with a magnitude of $R= 23.67$ (Fadda et al.
2004). The optical redshift of this source is unknown. Using the $R$
magnitude-redshift relationship for radio sources given by Rigby, Snellen, \&
Best [2007; their eqs.~(9) and (10)] we find $z\sim 1.55$. Within the rather
large uncertainties associated to this relationship, especially at $z \simgt 1$
(see Fig. 6 of Rigby et al. 2007), this estimate is compatible with the radio
source being at the estimated redshift of our {\it Spitzer} sources. If located
at $z=1.73$, the total luminosity at 1.4 GHz for the ``concordance'' cosmology
(spatially flat universe with $\Omega_\Lambda=0.7$  and
$H_0=70\,\hbox{km}\,\hbox{s}^{-1}\,\hbox{Mpc}^{-1}$), and a spectral index
$\alpha = 1$, is $7\cdot 10^{33}\,\hbox{erg}\,\hbox{s}^{-1}\,\hbox{Hz}^{-1}$,
i.e. a luminosity  which is typical of an FR II (Fanaroff \& Riley 1974) radio galaxy,
consistent with its edge-brightened morphology.

The total number of radio sources in the VLA-FLS area, as obtained by Condon et al (2003), 
brighter than 30 mJy is 18. We then set, following Brookes et al. (2008), the fraction of radio sources
at z$>$2 to be about 10\%. By doing this we expect about 2 extended radio sources in
the VLA-FLS field brighter than 30 mJy and with z$>$2. 
This number coincides with the two examples of extended radio sources associated to
overdensities of Spitzer-obscured galaxies in the FLS (the other case is currently under study).
We can then conclude that, despite of the small statistics there is a possible association
of extended radio sources at $z \sim 2$ with active mid-IR galaxies.


\section{Conclusions}

To test the possibility that the selected sources belong to a candidate
protocluster and shed light on the nature of optically obscured
galaxies defining the candidate filaments, we have carried out observations with
the MAMBO array at the IRAM 30m antenna on Pico Veleta of five optically
obscured sources. These latter trace one of the best examples of a candidate filament,
centered on a radio galaxy and roughly aligned with its jets. One of the {\it
Spitzer} sources is located right at the edge of one of its radio lobes. Four
of the observed sources were detected.
Although the available data are
insufficient to draw firm conclusions, the SEDs, obtained by combining our
measurements with optical, infrared and radio observations, are consistent with
the 4 detected sources being at the same $z\simeq 1.73$ redshift and,
therefore, tracing a real filament with a physical size of $\simeq 4\,$Mpc,
which may evolve into a rich galaxy cluster centered on a powerful
radio-galaxy.

The 4 detected sources most likely lie
in the region of the [${24\mu\hbox{m}}/{8\mu\hbox{m}}$,
${8\mu\hbox{m}}/{4.5\mu\hbox{m}}$] colour-colour plot  occupied by dust
obscured galaxies (DOGs; Pope et al. 2008). The Arp220 SED provides
reasonably acceptable fits for them. At the estimated redshift, the fits
imply bolometric luminosities in the range 5--$7\cdot 10^{12}$ L$_\odot$
and star formation rates of $\sim 1000$ M$_\odot$/yr. Their IR colours,
however, are bluer than those of sub-mm bright galaxies, suggesting either
a significant AGN contribution to IRAC fluxes or the presence of warmer
dust, perhaps heated by young starbursts. In fact, adding an obscured AGN
component improves the $\chi2$ values for three of the 4 sources. The
best fit AGN contributions to the bolometric luminosities range from 14 to
26\% of the total. This is enough for the AGN to dominate the emission at
5.8, 8 and 24$\mu$m, while the SB dominates at 1.25mm and at $\lambda
<4.5\mu$m. The new fits are again consistent with all these sources being
at the same redshift, $\simeq 1.73$.

The source J172030.9+592310 may be a foreground object at $z\simeq 1.2$,
although the redshift estimate is very uncertain.  Its SED is best interpreted
as the combination of that of Arp220 with the ``torus template'', corresponding
to the model used to fit the SED of a heavily obscured type 2 QSO, SWIRE
J104409.95+585224.8 (Polletta et al. 2006). At $z\simeq 1.2$, the best fit
luminosities of the starburst and of the AGN components are
$1.0\times  10^{12}\,L_\odot$ and $0.55\times 10^{12}\,L_\odot$, respectively.

A large-scale interplay between AGN activity (such as that produced by radio
jets) and star formation in the sources belonging to the protocluster may be at
work here and may shape the properties of these sources.\hfill\break
Following West (1994) we suggest that the protocluster may be in the process
of forming the dominant
cluster galaxy from the central powerful radio galaxy.
\hfill\break
To substantiate this
rather speculative interpretation we are extending this investigation to other
similar fields.

\section{Acknowledgements}

This work is based in part on observations taken with the IRAM 30m antenna,
IRAM is supported by INSU/CNRS (France), MPG (Germany) and IGN (Spain). We are
grateful to the IRAM staff at Pico Veleta and in particular to Stephan Leon for
his assistance with the pool observations. We also thank A. Bressan for having
made available the galaxy templates of Vega et al. (2008) in tabular form. GDZ
acknowledges partial financial support from ASI contracts I/016/07/0 ``COFIS''
and ``Planck LFI Activity of Phase E2''.
The authors wish to thank the referee for her/his suggestions that
helped in improving the paper.

{}

\clearpage

\end{document}